\documentclass[%
groupedaddress,
reprint,
showpacs,
 amsmath,amssymb,
 aps,
prb,
]{revtex4-1}

\newcommand\beq{\begin{equation}}
\newcommand\eeq{\end{equation}}

\usepackage{graphicx}
\usepackage{dcolumn}
\usepackage{bm}

\begin{document}


\title{ \boldmath$\mathcal{PT}$-Symmetry-Induced Wave Confinement and Guiding in Epsilon-Near-Zero Metamaterials}

\author{Silvio Savoia}
\author{Giuseppe Castaldi}
\author{Vincenzo Galdi}
\email{vgaldi@unisannio.it}
\affiliation{Waves Group, Department of Engineering, University of Sannio, I-82100 Benevento, Italy
}%

\author{Andrea Al\`u}
\affiliation{Department of Electrical and Computer Engineering, The University of Texas at Austin, Austin, TX 78712, USA}%

\author{Nader Engheta}
\affiliation{Department of Electrical and Systems Engineering, University of Pennsylvania, Philadelphia, PA 19104, USA}%

\date{\today}


\begin{abstract}
Inspired by the {\em parity-time} symmetry concept, we show that a judicious spatial modulation of gain and loss in {\em epsilon-near-zero} metamaterials can induce the propagation of exponentially-bound interface modes characterized by zero attenuation. With specific reference to a bi-layer configuration, via analytical studies and parameterization of the dispersion equation, we show that this waveguiding mechanism can be sustained in the presence of moderate gain/loss levels, and it becomes {\em leaky} (i.e., radiative) below a gain/loss threshold. Moreover, we explore a possible rod-based metamaterial implementation, based on realistic material constituents, which captures the essential features of the waveguiding mechanism, in good agreement with our theoretical predictions. Our results may open up new possibilities for the design of optical devices and reconfigurable nanophotonics platforms.
\end{abstract}


\maketitle

\section{Introduction}

The possibility to spatially modulate loss and gain brings about new dimensionalities in the design of metamaterials, which extend far beyond traditional loss-compensation schemes. Within this framework, particularly inspiring is the concept of {\em parity-time} ($\mathcal{PT}$) symmetry, originally conceived in quantum physics.\cite{Bender:1998,Bender:2002,Bender:2007}  Against the standard assumptions in quantum mechanics, Bender and co-workers \cite{Bender:1998,Bender:2002,Bender:2007} proposed an extended theory where the
Hermitian property of the Hamiltonian was replaced by a weaker symmetry condition on the quantum potential, $V(x) = V^*(-x)$,
involving the combined {\em parity} (i.e., spatial reflection, $\mathcal{P}$) and {\em time-reversal} (i.e., complex-conjugation, $\mathcal{T}$)
operator. They showed that, albeit {\em non-Hermitian}, such $\mathcal{PT}$-symmetric systems may still exhibit entirely real eigenspectra provided that their eigenstates are likewise $\mathcal{PT}$-symmetric. However, in view of the antilinear character of the $\mathcal{PT}$ operator, this last condition may not hold beyond some non-Hermiticity threshold, and the system may undergo a ``spontaneous symmetry breaking'', i.e., an abrupt phase transition to a complex eigenspectrum.\cite{Bender:1998,Bender:2002,Bender:2007}

In view of the formal analogies between quantum mechanics and (paraxial) optics, such concept can be translated to electromagnetic structures by means of spatial modulation of loss and gain, which is becoming technologically viable. 
In particular, optical ``testbeds'' of $\mathcal{PT}$-symmetric Hamiltonians have been proposed,\cite{Ruschhaupt:2005,El-Ganainy:2007} 
and experimentally characterized in either passive\cite{Guo:2009} (pseudo-$\mathcal{PT}$-symmetric) and actual gain-loss\cite{Ruter:2010}
configurations. Moreover, a variety of $\mathcal{PT}$-symmetry-inspired exotic effects have been observed in
optical, plasmonic, circuit-based, and metamaterial structures, including unidirectional propagation phenomena (invisibility, tunneling, negative refraction), coherent perfect absorption, beam switching, and absorption-enhanced transmission, with very promising potential applications to novel photonic devices and components (see Refs.
\onlinecite{Makris:2008,Longhi:2009ks,
Longhi:2010co,Ctyroky:2010de,
Chong:2011ev,Benisty:2011jr,Regensburger:2012,
Schindler:2012,Lin:2012,Ge:2012bq,
Zhu:2013jf,Feng:2013,Luo:2013,Kang:2013dg,
Castaldi:2013,Regensburger:2013kd,Savoia:2014,Peng:2014,Alaeian:2014eb,Alaeian:2014dj,Sun:2014fy,Feng:2014ve,Longhi:2014bn,
Fleury:2014iv,Ge:2014fd}
 for a sparse sampling). More recently, potential applications have also been proposed in connection with magnetic\cite{Lee:2014vd} and acoustic\cite{Zhang:2014bh} structures.
It is worth stressing that the potential technical issues that have been recently pointed out\cite{Lee:2014} in connection with the $\mathcal{PT}$-symmetric extension of quantum mechanics do not affect these electromagnetic and acoustic analogues.

In this paper, we present a study of $\mathcal{PT}$-symmetry-induced waveguiding in metamaterial slabs. Wave propagation at an interface between two media typically requires one of them to be conducting or with a negative real part of the permittivity. Here, on the contrary,
we study the propagation of exponentially-bound modes that can be sustained at a gain-loss interface under $\mathcal{PT}$-symmetry conditions, without requiring any modulation of the real part of permittivity across the interface. This intriguing propagation mechanism does not require negative values of the permittivity (real-part), and it is characterized by a purely real propagation constant (i.e., no attenuation). However, in order to achieve substantial localization along the transverse direction, {\em unfeasibly high} values of gain are generally required.\cite{Ctyroky:2010de}
We therefore suggest to operate in the {\em epsilon near zero} (ENZ) regime\cite{Engheta:2013il} (i.e., vanishingly small real part of the permittivities), in view of its well-known capabilities to dramatically enhance the effects of relatively low levels of loss and/or gain.\cite{Jin:2011kd,Sun:2012gq,Feng:2012}

Accordingly, the rest of the paper is laid out as follows. In Sec. \ref{Sec:Problem}, we introduce the waveguiding mechanism and discuss its attractive features as well as its limitations. In Sec. \ref{Sec:ENZ}, with specific reference to the ENZ regime, we analytically derive the dispersion equation for a $\mathcal{PT}$-symmetric bi-layer, and we identify a threshold condition on the gain/loss level which separates the bound- and leaky-mode regions. In Sec. \ref{Sec:Implementation}, we explore a possible rod-based metamaterial implementation which relies on a realistic (semiconductor) gain material.
Finally, in Sec. \ref{Sec:Conclusions}, we provide some concluding remarks and perspectives.

%
\begin{figure}
\begin{center}
\includegraphics [width=8cm]{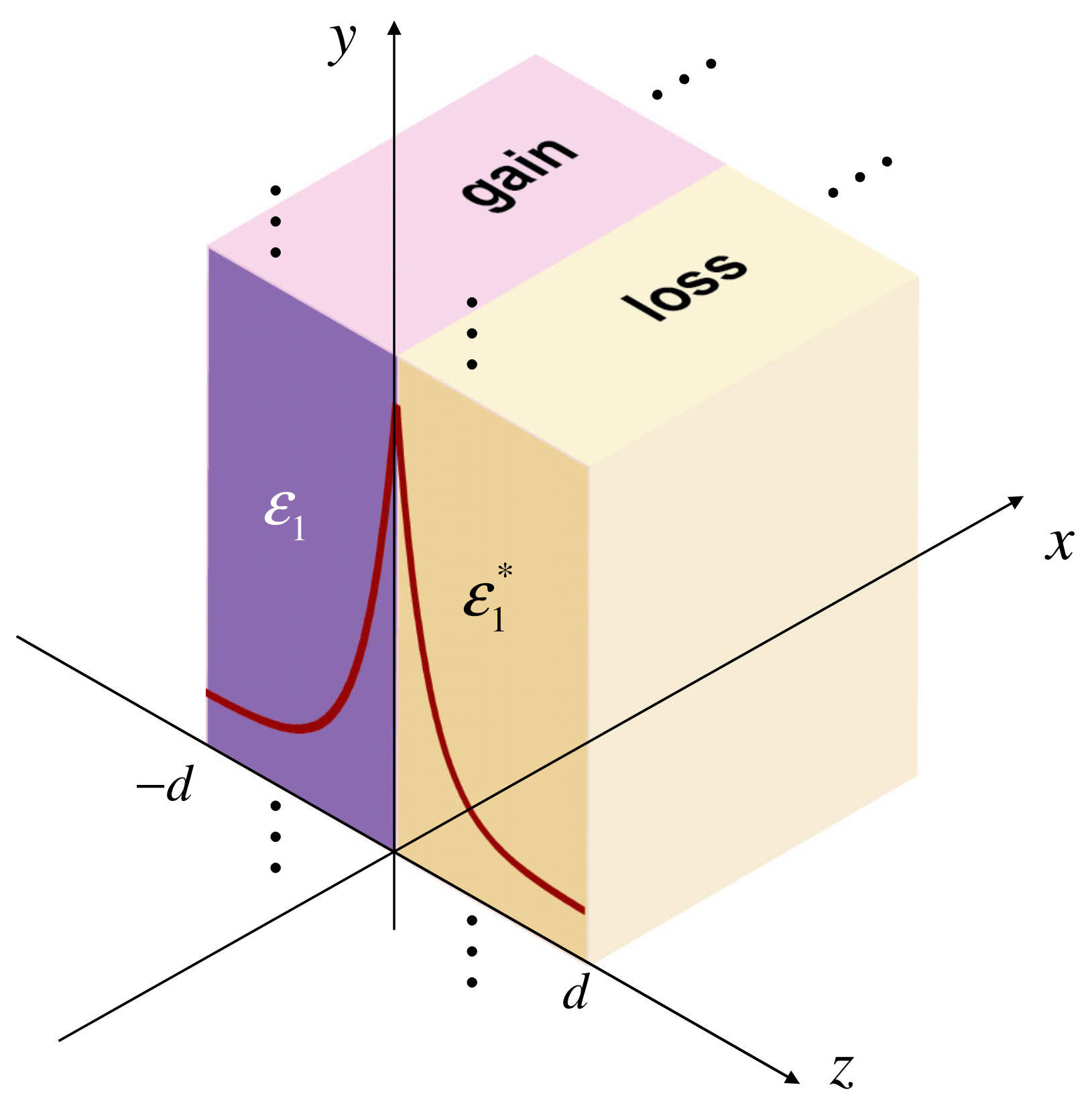}
\end{center}
\caption{(Color online) Problem schematic. A $\mathcal{PT}$-symmetric bi-layer consisting of two slabs of identical thickness $d$, and relative permittivity distribution as in (\ref{eq:epss}), which can support TM-polarized modes exponentially bound and the gain-loss interface $z=0$, and propagating without attenuation along the $x$ direction.}
\label{Figure1}
\end{figure}

\section{Background and Problem Statement}
\label{Sec:Problem}

\subsection{Geometry}

With reference to the schematic in Fig. \ref{Figure1}, we start considering an isotropic, non-magnetic, piece-wise homogeneous bi-layer composed of two slabs of identical thickness $d$, immersed in vacuum, infinitely extent in the $x,y$ plane, and paired along the $z$-direction. Our model is hence parameterized by the relative permittivity distribution
\beq
\varepsilon\left(z\right)=\left\{
\begin{array}{lll}
1,\hspace{6mm}\left|z\right|>d,\\
\varepsilon_1,\hspace{5mm} -d<z<0,\\
\varepsilon_1^*,\hspace{5mm} 0<z<d,
\end{array}
\right.
\label{eq:epss}
\eeq
where $\varepsilon_1=\varepsilon'-i\varepsilon''$, with $\varepsilon'>0$, $\varepsilon''>0$. Under the assumed time-harmonic [$\exp(-i\omega t)$] convention, this implies that the regions $-d<z<0$ and $0<z<d$
are characterized by gain and loss, respectively,  and the structure fulfills the necessary condition for $\mathcal{PT}$ symmetry,
\beq
\varepsilon\left(
z
\right)=\varepsilon^*\left(-z\right).
\eeq

\subsection{$\mathcal{PT}$-symmetry-induced surface-waves}
In Ref. \onlinecite{Ctyroky:2010de}, it was pointed out that, for transverse-magnetic (TM) polarization (i.e., $y$-directed magnetic field), the structure in Fig. \ref{Figure1} may support a $\mathcal{PT}$-induced surface wave exponentially bound at the gain-loss interface $z=0$. This waveguiding mechanism is perhaps more easily understood in the half-space limit $d\rightarrow\infty$, for which the dispersion relationship is simply given by (see Appendix \ref{Sec:AppA} for details)
\beq
k_x=k_0\sqrt{\frac{\varepsilon_1\varepsilon_1^*}{\varepsilon_1+\varepsilon_1^*}}=k_0\sqrt{\frac{\left(\varepsilon'\right)^2+\left(\varepsilon''\right)^2}{2\varepsilon'}},
\label{eq:kSW}
\eeq
with $k_0=\omega/c_0=2\pi/\lambda_0$ denoting the vacuum wavenumber (and $c_0$ and $\lambda_0$ the corresponding wavespeed and wavelength, respectively). Accordingly, the field localization in the gain and loss regions is controlled by the (complex) transverse wavenumbers
\beq
k_{z1}=\sqrt{\varepsilon_1k_0^2-k_x^2},~~\mbox{Im}\left(k_{z1}\right)\le0,
\label{eq:kz1}
\eeq
and $k_{z1}^*$, respectively.

The dispersion relationship in (\ref{eq:kSW}) can be interpreted as a generalization of the Zenneck-wave\cite{Zenneck:1907} and surface-plasmon-polariton\cite{Maier:2007} (SPP) cases, featuring oppositely signed
imaginary parts of the permittivities. By comparison with these two latter cases, the following observations are in order:
\begin{itemize}
\item[{\em i)}]{Both media exhibit the same {\em positive} value of permittivity (real-part), and therefore the mechanism differs substantially from gain-assisted SPP-propagation schemes.\cite{Nezhad:2004fg}}
\item[{\em ii)}]{The $\mathcal{PT}$-symmetry condition inherently yields a {\em real-valued} propagation constant $k_x$, i.e., {\em unattenuated} propagation along the gain-loss interface.}
\item[{\em iii)}]{From the physical viewpoint, such waveguiding mechanism is sustained by a transverse (i.e., $z$-directed) component of the power flux from the gain- to the loss-region.}
\item[{\em iv)}]{The branch-cut choice in the gain region [cf. (\ref{eq:kz1})] may appear somewhat arbitrary, given that the usual radiation condition and decay at infinity cannot be used as an argument in a gain background.
Indeed, this a rather controversial issue in the literature (see, e.g., Refs. \onlinecite{Skaar:2006yh,Nazarov:2007hl,Nazarov:2007lp,Govyadinov:2007uh,Lakhtakia:2007sf,Siegman:2010lj} for a sparse sampling). We point out, however, that this choice is {\em irrelevant} for the bi-layer scenario of actual interest here, and it only matters for the half-space configuration.\cite{Lakhtakia:2007sf} This latter is, however, an unrealistic limit that we consider only in view of the particularly simple form of the dispersion relationship. Nevertheless, for several representative values of $\varepsilon'$ and $\varepsilon''$ (within and beyond the ENZ regime), we verified numerically that the choice in (\ref{eq:kz1}) yields results that are consistently in agreement with those obtained by truncating (along $z$) the half-space configuration at distances for which the field is sufficiently decayed.}
\end{itemize}
The above waveguiding mechanism looks potentially attractive under many respects. For instance, one may envision nanophotonics platforms where channels of gain media are suitably embedded in a lossy background, so that the waveguiding may be selectively enabled (and possibly reconfigured) by optically pumping certain spatial regions. So, effectively we may have ``waveguiding on demand", where and when we want it.
This may bring about new perspectives and degrees of freedom in the design of optical switches, modulators, and reconfigurable photonic networks.

%
\begin{figure}
\begin{center}
\includegraphics [width=8cm]{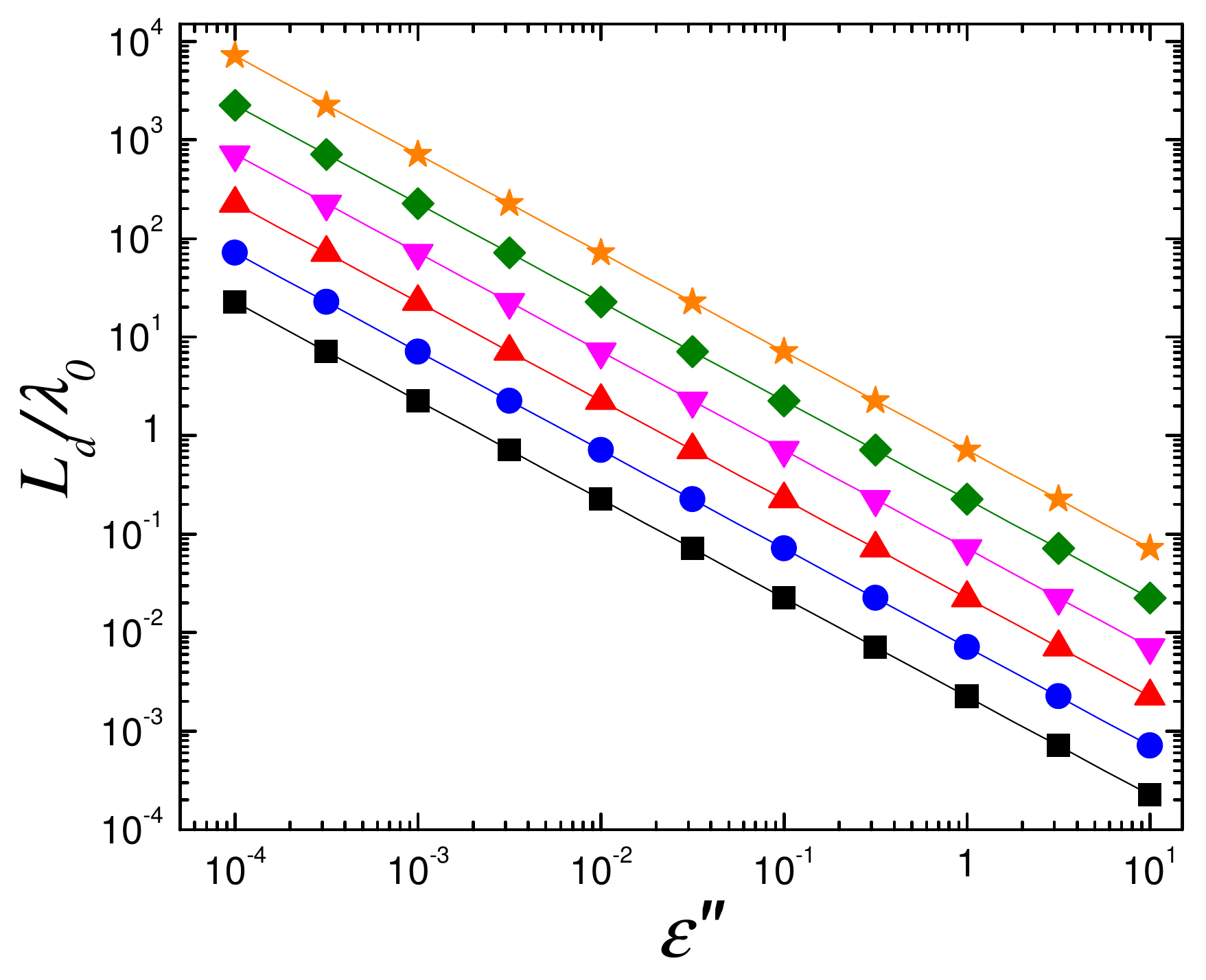}
\end{center}
\caption{(Color online) Geometry as in Fig. \ref{Figure1}, but in the asymptotic limit $d\rightarrow \infty$. Decay length [cf. (\ref{eq:Ld})], scaled by the vacuum wavelength, as a function of the gain/loss level $\varepsilon''$, for various representative values of the relative-permittivity real part: $\varepsilon'=10^{-4}$ (squares), $\varepsilon'=10^{-3}$ (circles), $\varepsilon'=10^{-2}$ (up-triangles), $\varepsilon'=0.1$ (down-triangles), $\varepsilon'=1$ (diamonds), $\varepsilon'=10$ (stars).}
\label{Figure2}
\end{figure}

\subsection{Transverse localization vs. gain/loss level}
Although, in view of (\ref{eq:kz1}), the half-space limit always features exponential decay (along $z$) of the fields, one intuitively expects the localization to depend critically on the  gain/loss level (and to vanish in the absence of gain and loss). For a more quantitative assessment of such localization properties, we show in Fig. \ref{Figure2} the decay length\cite{Maier:2007}
\beq
L_d=\frac{1}{\left|\mbox{Im}\left(k_{z1}\right)\right|},
\label{eq:Ld}
\eeq
as a function of the imaginary part (absolute value) of the permittivity $\varepsilon''$, for representative values of the real part $\varepsilon'$ spanning several orders of magnitude. As evidenced by the log-log scale, for a given wavelength and relative-permittivity real-part, the decay length decreases algebraically with increasing values of the gain/loss level. In particular, localization on subwavelength scales requires values of $\varepsilon''$ that are of the same order or even larger than $\varepsilon'$.
Thus, assuming for instance $\varepsilon'=10$ (compatible with semiconductor materials at optical wavelengths), gain/loss levels as high as $\varepsilon''=3$ would be required to attain a decay length $L_d\sim \lambda_0/4$. To give an idea, at the telecom wavelength $\lambda_0=1550$ nm, this corresponds to a gain coefficient $\gamma=4\pi\mbox{Im}\left(\sqrt{\varepsilon_1}\right)/\lambda_0\sim 38000$ cm$^{-1}$, i.e., about an order of magnitude larger than the largest gain levels attainable with current technologies.\cite{Hatori:2000,Nezhad:2004fg,Yu:2008,Geskus:2012}

What also clearly emerges from Fig. \ref{Figure2} is that decreasing $\varepsilon'$ may allow working with substantially lower gain/loss levels. For instance, assuming $\varepsilon'=10^{-4}$, decay lengths $L_d\sim \lambda_0/4$ could be attained with gain/loss levels $\varepsilon''\sim 0.009$, i.e., gain coefficients (at $\lambda_0=1550$nm) $\gamma\sim 5000$ cm$^{-1}$.

\section{The ENZ Regime}
\label{Sec:ENZ}
From the above results and observations, it turns out that the ENZ regime, 
\beq
\varepsilon'\ll \varepsilon''\ll 1,
\label{eq:ENZ1}
\eeq
seems particularly promising for the waveguiding mechanism of interest. While the desired ENZ $\mathcal{PT}$-symmetric characteristics 
cannot be found in natural materials, we show hereafter (see Sec. \ref{Sec:Implementation} below) that they can be artificially engineered based on realistic material constituents. Before that, however, we study in detail the more realistic bi-layer (i.e., finite $d$) scenario in Fig. \ref{Figure1} in the ENZ regime (\ref{eq:ENZ1}).

\subsection{Dispersion equation: Bound vs. leaky modes}

It can be shown (see Appendix \ref{Sec:AppB} for details) that a $\mathcal{PT}$-symmetric ENZ metamaterial bi-layer [cf. Fig. \ref{Figure1}] supports modes propagating along the $x$ direction with a {\em generally complex} propagation constant $k_x$ which
satisfies the dispersion equation
\beq
ik_{z0}\lbrace|\tau_1|^2\mbox{Re}[\varepsilon_1^2(k_{z1}^*)^2]-|\varepsilon_1|^2|k_{z1}^2|
\rbrace
-|k_{z1}|^2 \mbox{Re}(\varepsilon_1k_{z1}^*\tau_1^*)=0,
\label{eq:DD}
\eeq
where
\beq
k_{z0}=\sqrt{k_0^2-k_x^2},
\label{eq:kz0}
\eeq
and
\beq
\tau_1=\tan\left(k_{z1}d\right).
\label{eq:tau1}
\eeq
In view of the inherent geometrical symmetry, without loss of generality, we focus hereafter on the case $\mbox{Re}(k_x)>0$ (i.e., propagation along the positive $x$ direction).   
Among the possible solutions of (\ref{eq:DD}) in the complex $k_x$ plane, we are especially interested in {\em bound} modes characterized by
\beq
\mbox{Re}\left(k_x\right)>k_0,~~\mbox{Im}\left(k_{z0}\right)\ge 0,
\label{eq:BW}
\eeq 
i.e., an exponential decay in the exterior vacuum region $|z|>d$. While it is well-known that no such mode can be sustained by a low-permittivity slab in the absence of loss and gain, it can be shown (see Appendix \ref{Sec:AppC} for details) that 
this becomes possible for gain/loss levels beyond a threshold value 
\beq
\varepsilon^{\prime\prime}_t=\sqrt{\frac{\varepsilon^\prime(2-\varepsilon^\prime)[\varepsilon^\prime k_0d(\tau_0^2-1)+2\tau_0]}{k_0d(\varepsilon^\prime-2)(\tau_0^2-1)+2\tau_0}},~~\tau_0=\tanh\left(k_0d\right),
\label{eq:varepsilonc}
\eeq
and it also implies $\mbox{Im}(k_x)=0$ (i.e., no attenuation).  
Below such threshold, {\em leaky} modes can instead be found, characterized by complex propagation constants
\beq
\mbox{Re}\left(k_x\right)<k_0,~~\mbox{Im}\left(k_x\right)>0,~~\mbox{Im}\left(k_{z0}\right)\le 0.
\label{eq:LW}
\eeq
To avoid possible confusion, we stress that the complex character of these latter solutions is by no means related to the aforementioned spontaneous symmetry breaking phenomenon in $\mathcal{PT}$-symmetric systems,\cite{Bender:1998,Bender:2002,Bender:2007} as it would also arise in the absence of gain and loss.\cite{Bahl:1974} 
These solutions exhibit exponential decay along the propagation direction $x$, and exponential growth along the transverse direction $z$. Although such character appears clearly unphysical, they have long been utilized in the antenna community to effectively model physical resonant radiative states in waveguides.\cite{Jackson:1988}

%
\begin{figure}
\begin{center}
\includegraphics [width=8.5cm]{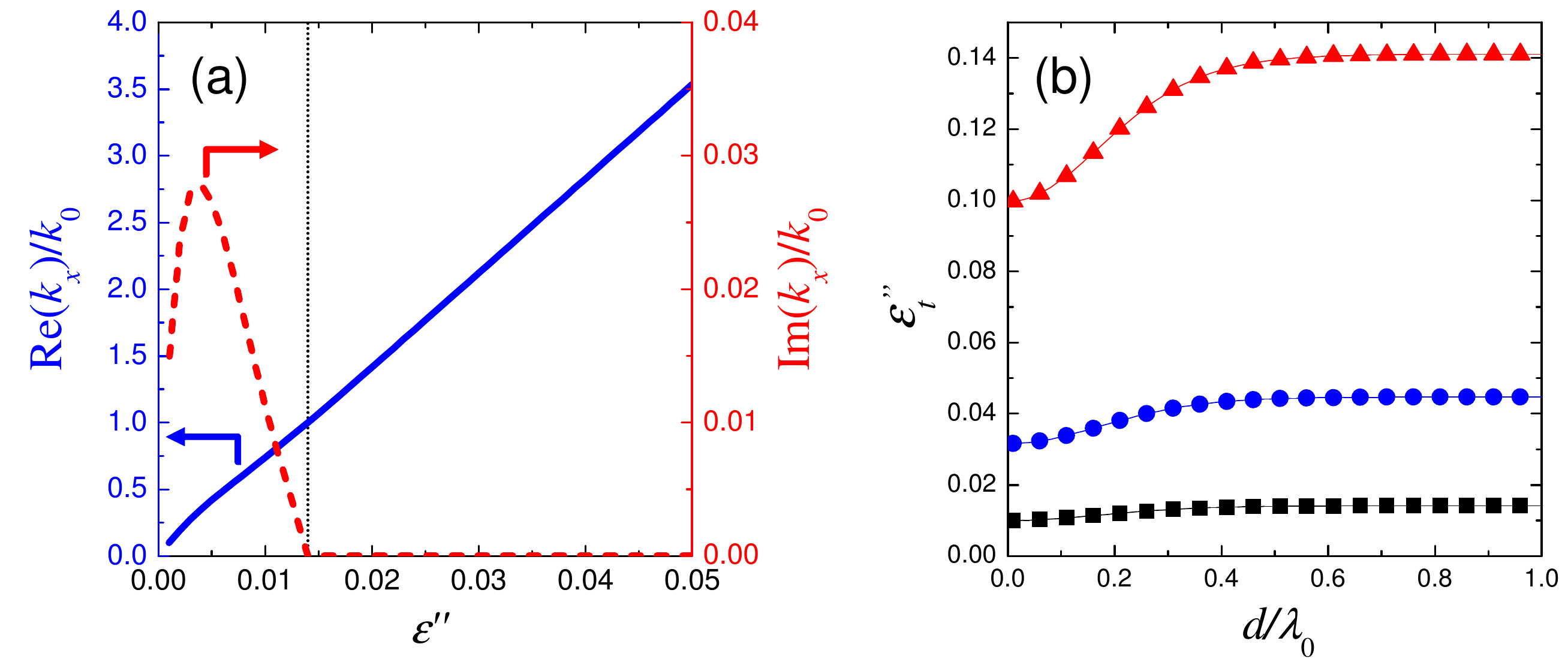}
\end{center}
\caption{(Color online) Geometry as in Fig. \ref{Figure1}, for $\varepsilon'=10^{-4}$ and $d=0.5\lambda_0$. (a) Real- (blue-solid; left axis) and imaginary-part (red-dashed; right axis) of the numerically-computed [from (\ref{eq:DD})] propagation constant, as a function of the gain/loss level $\varepsilon''$, illustrating the transition from leaky to bound modes occurring at the threshold $\varepsilon''_t=0.014$ (black-dotted vertical line). 
 (b) Gain/loss level threshold [cf. (\ref{eq:varepsilonc})] as a function of $d/\lambda_0$, for  $\varepsilon'=10^{-4}$ (squares), $\varepsilon'=10^{-3}$ (circles), $\varepsilon'=10^{-2}$ (triangles).}
\label{Figure3}
\end{figure}

To illustrate the threshold phenomenon, Fig. \ref{Figure3}(a) shows the numerically-computed  propagation constant $k_x$ as a function of $\varepsilon''$, for given values of $\varepsilon'$ and the bi-layer electrical thickness. As it can be observed, for increasing values of $\varepsilon''$ there is a smooth transition from a leaky [cf. (\ref{eq:LW})] to a bound [cf. (\ref{eq:BW})] mode solution. The separation between these two regions occurs at the grazing condition $k_x=k_0$, and the corresponding gain/loss level is in very good agreement with the analytical estimate of the threshold $\varepsilon''_t$ in (\ref{eq:varepsilonc}). 

Figure \ref{Figure3}(b) shows the behavior of such threshold as a function of the bi-layer electrical thickness, for representative values of $\varepsilon'$. We observe that the threshold depends only mildly on the bi-layer electrical thickness and, for sufficiently thick bi-layers ($k_0d\gg 1$, i.e., $\tau_0\approx 1$), it approaches the asymptotic value
\beq
\varepsilon''_{t\infty}=\sqrt{\varepsilon'\left(2-\varepsilon'\right)}, 
\label{eq:asthresh}
\eeq 
which is consistent with enforcing $|k_x|>k_0$ in the asymptotic dispersion relationship (\ref{eq:kSW}). Moreover, as it can be expected, the threshold increases with increasing values of $\varepsilon'$, but maintains moderately small values within the ENZ regime of interest. We stress that the threshold in (\ref{eq:varepsilonc}) and its asymptotic limit in (\ref{eq:asthresh}) are only valid in the ENZ limit (\ref{eq:ENZ1}). Therefore, the fact that $\varepsilon''_{t\infty}$ in (\ref{eq:asthresh}) vanishes for $\varepsilon'=2$, and it becomes imaginary for $\varepsilon'>2$, by no means indicates that the threshold disappears for sufficiently thick bi-layers, but rather than the ENZ approximation is no longer valid in those parameter ranges.

%
\begin{figure*}
\begin{center}
\includegraphics [width=16cm]{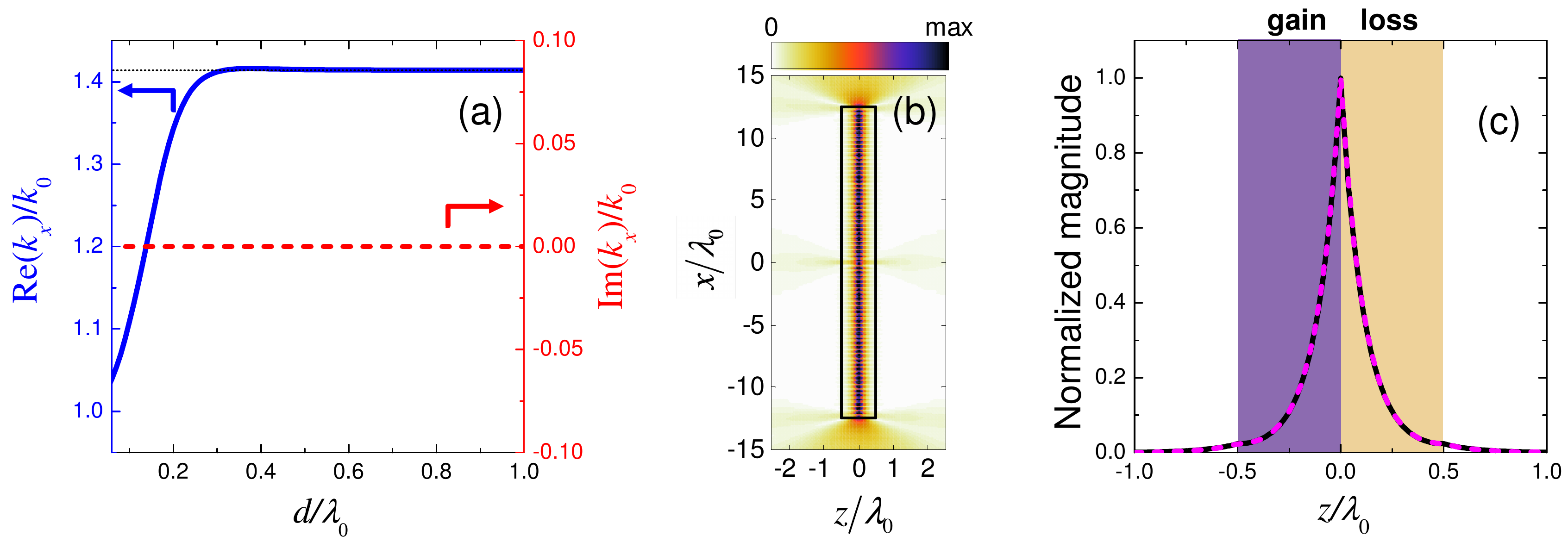}
\end{center}
\caption{(Color online) (a) As in Fig. \ref{Figure3}(a), but as a function of $d/\lambda_0$ (dispersion relationship), for $\varepsilon'=10^{-4}$ and $\varepsilon''=0.02$ (above-threshold case). Also shown (black-dotted horizontal line), as a reference, is the asymptotic limit (\ref{eq:kSW}).
(b) Numerically-computed field magnitude ($|H_y|$) map for a bi-layer with $d=0.5\lambda_0$ and finite-size (along $x$) width of $25\lambda_0$ (delimited by a black-solid rectangle), excited by a magnetic line source located at $x=z=0$. Values are sampled so as to avoid the singularity at the source, and are normalized with respect to the maximum. (c) Transverse cut (magenta-dashed) at $x=4.17\lambda_0$, compared with analytical bound-mode prediction [black-solid; cf. (\ref{eq:ee})] with $k_x=1.414k_0$.}
\label{Figure4}
\end{figure*}

%
\begin{figure*}
\begin{center}
\includegraphics [width=16cm]{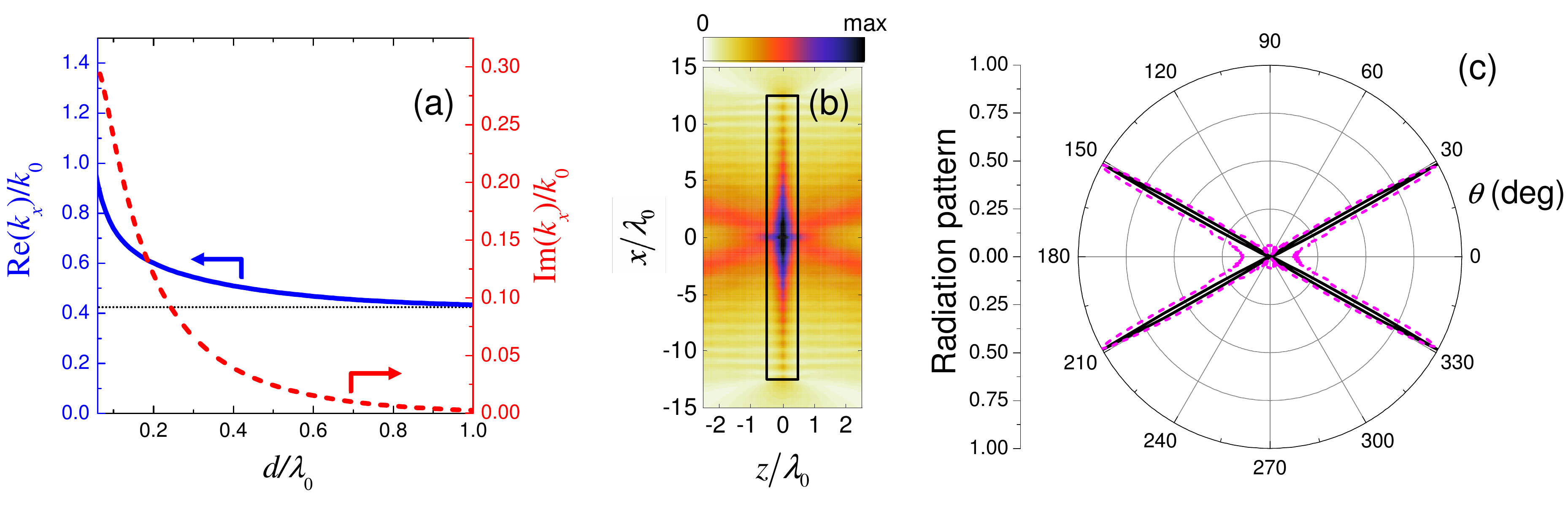}
\end{center}
\caption{(Color online) (a), (b) As in Figs. \ref{Figure4}(a) and \ref{Figure4}(b), respectively, but for $\varepsilon''=0.006$ (subthreshold case). (c) Numerically-computed radiation pattern (with the angle $\theta$ measured with respect to the $z$ axis) compared with leaky-mode-based theoretical prediction in (\ref{eq:LWRP}) for $k_x=(0.486+i0.02) k_0$.}
\label{Figure5}
\end{figure*}

\subsection{Representative results}
Figures \ref{Figure4} and \ref{Figure5} illustrate some representative results for $\varepsilon'=10^{-4}$ and two feasible gain/loss levels. More specifically, for an above-threshold case [$\varepsilon''=0.02$, cf. Fig. \ref{Figure3}(b)], Fig. \ref{Figure4}(a) shows the numerically-computed dispersion relationship of a bound mode. As theoretically predicted, we observe a {\em purely real} propagation constant (i.e., no attenuation), which approaches the asymptotic prediction [cf. (\ref{eq:kSW})] for $d/\lambda_0\gtrsim 0.3$. To verify the physical character of this mode and its actual excitability, Fig. \ref{Figure4}(b) shows a numerically-computed (see Appendix \ref{Sec:AppD} for details) near-field map pertaining to a finite-size (along $x$) structure excited by a magnetic line-source located at the gain-loss interface at $x=0$.
A bound-mode structure is clearly visible, with a standing-wave pattern originating from the structure truncation along the $x$ direction. For a more quantitative assessment, Fig. \ref{Figure4}(c) shows a transverse ($z$) cut, which clearly exhibits an exponential localization, and is in excellent agreement with the theoretical prediction (see Appendix \ref{Sec:AppB}).

Figures \ref{Figure5}(a)--\ref{Figure5}(c) illustrate the corresponding results for a subthreshold gain/loss level [$\varepsilon''=0.006$, cf. Fig. \ref{Figure3}(b)]. More specifically, in the dispersion relationship [Fig. \ref{Figure5}(a)] we now observe a {\em complex} propagation constant, which is indicative of a leaky mode [cf. (\ref{eq:LW})]. As also evident from the near-field map in Fig. \ref{Figure5}(b), this represents a physical resonant {\em radiative} state supported by the bi-layer. As a further confirmation, Fig. \ref{Figure5}(c) compares the numerically-computed  (far-field) radiation pattern with the theoretical leaky-mode-based prediction,\cite{Jackson:1988} 
\beq
\left|H_y^{(ff)}\right|^2\left(\theta\right)\sim A
\cos^2\theta\left\{
\frac{\alpha^2+\beta^2}{\left[k_0^2\sin^2\theta-\left(\beta^2-\alpha^2\right)\right]^2+\left(2\alpha\beta\right)^2}
\right\},
\label{eq:LWRP}
\eeq
where $k_x=\beta+i\alpha$ is the complex propagation constant [cf. (\ref{eq:LW})], $A$ is a normalization constant, and the angle $\theta$ is measured with respect to the $z$ axis. A good agreement is observed, with the discrepancies attributable to the finite-size aperture (along $x$) of the bi-layer.

%
\begin{figure*}
\begin{center}
\includegraphics [width=16cm]{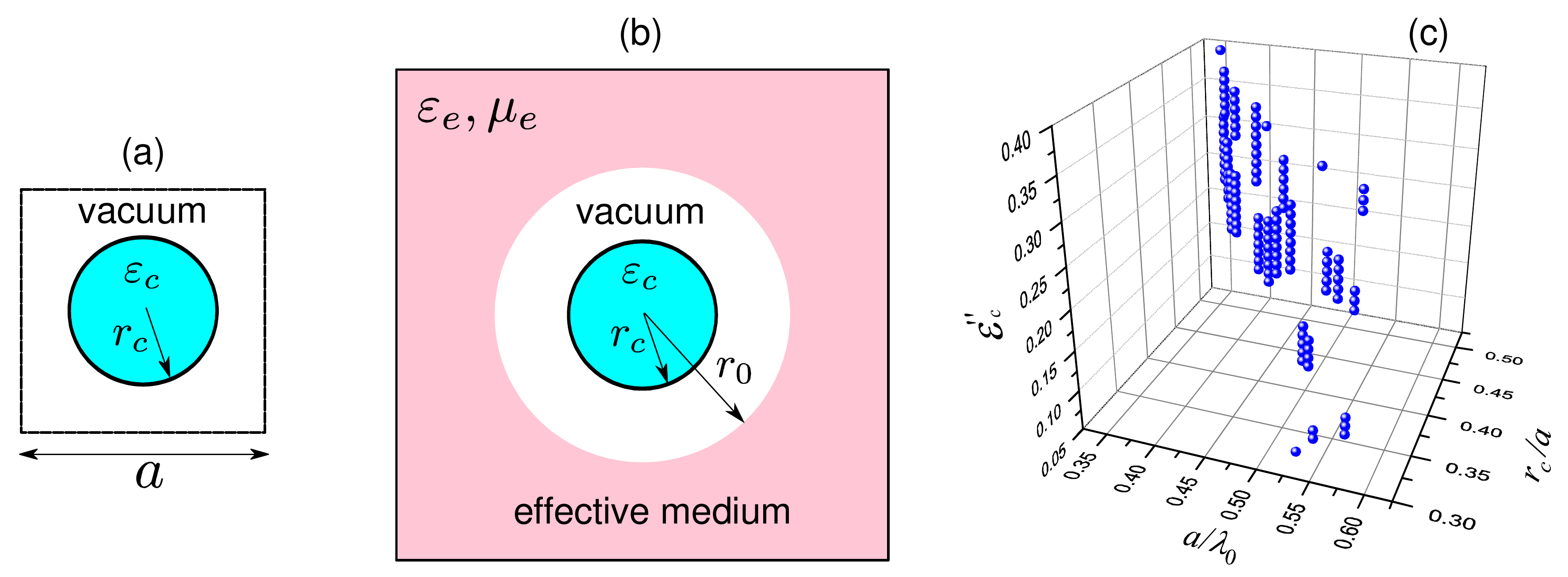}
\end{center}
\caption{(Color online) Metamaterial implementation. (a) Unit-cell describing a 2-D array of non-magnetic cylindrical rods of radius $r_c$ and relative permittivity $\varepsilon_c=\varepsilon'_c-i\varepsilon''_c$ arranged according to a square lattice with period $a$. (b) Schematics of the effective-medium model: vacuum-coated rod embedded in an effective medium of unknown parameters $\varepsilon_e$ and $\mu_e$. The radius $r_0$ is chosen so that the area of the coated rod is that of the actual square unit-cell in (a). (c) Representative results from the synthesis problem in the 3-D parameter space ($\varepsilon''_c, a/\lambda_0, r_c/a$), assuming $\varepsilon'_c=11.38$: each marker represents a candidate configuration  that satisfies the (asymptotic) condition in (\ref{eq:boundasmu}) for the existence of an unattenuated bound mode.}
\label{Figure6}
\end{figure*}

\subsection{Some remarks}
In essence, from a physical viewpoint, the threshold phenomenon implies that  
for low gain/loss levels, the transverse power flow from the gain to the loss region is not sufficient to sustain a bound mode, and the structure tends to radiate [at an angle and with a beam-width strictly related to the complex propagation constant,\cite{Jackson:1988} cf. (\ref{eq:LWRP})]. This is similar to what is observed in standard (lossless, gainless) low-permittivity slabs.\cite{Bahl:1974} 
By increasing the gain/loss level, the radiation direction progressively departs from the $z$ axis, and becomes grazing at the threshold value $\varepsilon''_t$ in (\ref{eq:varepsilonc}). Beyond this threshold, the transverse-power-flow mechanism becomes sufficiently effective for the structure to sustain a bound mode.

Incidentally, we found a similar threshold phenomenon (with identical parameterization) in a previous study\cite{Savoia:2014} dealing with the surface-wave-mediated {\em tunneling} of impinging waves through the same structure as in Fig. \ref{Figure1}. This is not surprising, based on reciprocity arguments.

Another interesting aspect of the above described waveguiding mechanism is that the propagation constant in the above-threshold (bound-mode) region is {\em inherently real}, irrespective of the gain/loss level and electrical thickness. In other words, these bound modes 
are not subject to the spontaneous symmetry breaking phenomenon that generally occurs in $\mathcal{PT}$-symmetric systems.\cite{Bender:1998,Bender:2002,Bender:2007} 
This is quite unusual, and not observable in other waveguiding mechanisms. To give an idea, for $\varepsilon'>1$, a $\mathcal{PT}$-symmetric bi-layer could also support higher-order guided modes, which may be viewed as the complex-valued transpositions of the standard guided modes supported by a dielectric (lossless, gainless) slab waveguide. For such modes, parameters could be tuned so as the propagation constant would stay real within certain ranges. However, by increasing the gain/loss level and/or the electrical thickness, spontaneous symmetry breaking would eventually occur, and the propagation constant would become complex.

\section{Possible Implementation}
\label{Sec:Implementation}
A typical implementation of ENZ metamaterials at optical wavelengths is based on multilayers combining thin subwavelength layers of positive- (e.g., dielectric) and negative-permittivity (e.g., metals or oxides) materials. In such implementations, the use of gain has been proposed in order to compensate the unavoidable loss effects.\cite{Rizza:2011be,Ni:2011} However, since an interface between a positive- and negative-permittivity material is naturally capable to support a surface wave also in the absence of balanced gain and loss, such implementation may not allow a clear-cut visualization and interpretation of the $\mathcal{PT}$-symmetry-induced waveguiding phenomenon of interest here. 

For a more effective illustration of our arguments, we therefore take inspiration from {\em all-dielectric} implementations of near-zero-refractive-index metamaterials based on periodic arrays of high-permittivity cylindrical rods exhibiting Dirac-cone dispersion at the center of the Brillouin zone.\cite{Huang:2011,Moitra:2013}

\subsection{Effective parameters}
As schematically illustrated in the unit-cell shown in Fig. \ref{Figure6}(a), we consider a possible implementation consisting of non-magnetic cylindrical rods of radius $r_c$ and relative permittivity $\varepsilon_c=\varepsilon'_c-i\varepsilon''_c$ arranged according to a square lattice with period $a$. As in Ref. \onlinecite{Huang:2011}, we model such metamaterial by means of the effective-medium theory developed in Ref. \onlinecite{Wu:2006hd}. In essence, as illustrated in Fig. \ref{Figure6}(b), such model assumes a vacuum-coated cylinder of total radius $r_0$ embedded in an effective medium of unknown parameters $\varepsilon_e$ and $\mu_e$. The radius $r_0$ is chosen so that the area of the coated cylinder is the one of the actual square unit-cell, and the effective parameters are computed by self-consistency, i.e., by enforcing that the total scattering of an electromagnetic wave vanishes. In particular, in the limit $k_er_0\ll 1$, we obtain\cite{Wu:2006hd} the simple equations\footnote{Note that our expressions differ from those in Ref. \citenum{Wu:2006hd} in view of the different polarization assumed.}
\beq
\frac{\varepsilon_e-\displaystyle{\frac{J_1\left(k_0r_0\right)}{k_0r_0J_1^\prime\left(k_0r_0\right)}}}{\varepsilon_e-\displaystyle{\frac{Y_1\left(k_0r_0\right)}{k_0r_0Y_1^\prime\left(k_0r_0\right)}}}=\frac{Y_1^\prime\left(k_0r_0\right)}{iJ_1^\prime\left(k_0r_0\right)}\left(\frac{D_1}{1+D_1}\right),
\label{eq:epse}
\eeq
\beq
\frac{\mu_e+\displaystyle{\frac{2J_0^\prime\left(k_0r_0\right)}{k_0r_0J_0\left(k_0r_0\right)}}}{\mu_e+\displaystyle{\frac{2Y_0^\prime\left(k_0r_0\right)}{k_0r_0Y_0\left(k_0r_0\right)}}}=\frac{Y_0\left(k_0r_0\right)}{iJ_0\left(k_0r_0\right)}\left(\frac{D_0}{1+D_0}\right),
\label{eq:mue}
\eeq
which can readily be solved analytically in closed form. In (\ref{eq:epse}) and (\ref{eq:mue}),
$k_c=k_0\sqrt{\varepsilon_c}$, $J_{\nu}$ and $Y_{\nu}$ are the $\nu$th-order Bessel and Neumann functions,\cite{Abramowitz:1964} respectively, the prime denotes differentiation with respect to the argument, and
\beq
D_{\nu}=\frac{k_c J_{\nu}^\prime\left(k_cr_c\right)J_{\nu}\left(k_0r_c\right)-\varepsilon_c k_0 J_{\nu}\left(k_cr_c\right)J_{\nu}^\prime\left(k_0r_c\right)}{\varepsilon_c k_0 J_{\nu}\left(k_cr_c\right)H_{\nu}^{\prime\left(1\right)}\left(k_0r_c\right)-k_c J_{\nu}^\prime\left(k_cr_c\right)H_{\nu}^{\left(1\right)}\left(k_0r_c\right)},
\eeq
with $H_{\nu}^{(1)}$ denoting the $\nu$th-order Hankel function of the first kind.\citep{Abramowitz:1964}
Referring to Ref. \citenum{Wu:2006hd} for a thorough assessment of the range of applicability of the above model, we stress that the underlying approximation does not require $k_0r_0$, $k_0 r_c$ and $k_c r_c$ to be small, and thus its validity can extend beyond the standard long-wavelength limit.

\subsection{Model generalizations}
In view of the generally magnetic character of the effective medium, our $\mathcal{PT}$-symmetric model in Fig. \ref{Figure1} needs to be generalized, by assuming also a relative permeability distribution
\beq
\mu\left(z\right)=\left\{
\begin{array}{lll}
1,\hspace{6mm}\left|z\right|>d,\\
\mu_1,\hspace{5mm} -d<z<0,\\
\mu_1^*,\hspace{5mm} 0<z<d,
\end{array}
\right.
\label{eq:mus}
\eeq
where $\mu_1=\mu'-i\mu''$, with $\mu'>0$, $\mu''>0$. Accordingly, the dispersion relationship of a TM-polarized bound mode in the asymptotic ($d\rightarrow\infty$) limit can be generalized as follows (see Appendix \ref{Sec:AppA} for details)
\begin{eqnarray}
k_x&=&
k_0
\sqrt{\frac{\varepsilon_1\varepsilon_1^*\left(\varepsilon_1\mu_1^*-\varepsilon_1^*\mu_1\right)}
{
\varepsilon_1^2-\left(
\varepsilon_1^*
\right)^2
}}\nonumber\\
&=& k_0\left|
\varepsilon_1\right|\sqrt{\frac{\varepsilon^{\prime\prime}\mu^\prime-\varepsilon^\prime\mu^{\prime\prime}}{2\varepsilon^{\prime\prime}\varepsilon^\prime}},
\label{eq:kxSW1}
\end{eqnarray}
subject to the further condition
\beq
\mbox{Im}\left(\frac{k_{z1}}{\varepsilon_1}\right)=0,
\label{eq:addcond}
\eeq
where
\beq
k_{z1}=\sqrt{k_0^2\varepsilon_1\mu_1-k_x^2},~~\mbox{Im}\left(k_{z1}\right)\le0.
\label{eq:kz1mu}
\eeq
For the bi-layer (i.e., finite $d$) case, the dispersion equation remains formally identical to (\ref{eq:DD}), but with $k_{z1}$ defined in (\ref{eq:kz1mu}). In principle, it is also possible to generalize the threshold condition in (\ref{eq:varepsilonc}), but the derivation is rather cumbersome. Instead, we consider the asymptotic limit $d/\lambda_0\gg 1$ (of direct interest for our subsequent studies), for which the existence of a bound mode can be established by enforcing in (\ref{eq:kxSW1}) real-valued solutions with $k_x> k_0$, which yields
\beq
\frac{\mu^{\prime\prime}}{\varepsilon^{\prime\prime}}<
\frac{\mu^\prime}{\varepsilon^\prime}-\frac{2}{\left(\varepsilon^{\prime\prime}\right)^2+\left(\varepsilon^{\prime}\right)^2}.
\label{eq:boundasmu}
\eeq

\subsection{Synthesis}
In view of the simple analytical structure of the effective-medium model in (\ref{eq:epse}) and (\ref{eq:mue}), and the limited number of parameters, we found it computationally effective to synthesize the metamaterial via a constrained parameter search. In what follows, we focus on the synthesis of the gain region, which entails $\varepsilon''_c>0$; it
is easily verified from (\ref{eq:epse}) and (\ref{eq:mue}) that the lossy counterpart can be obtained by
changing the sign of $\varepsilon_c''$.

In our synthesis, we fix the real part of the relative permittivity of the rods $\varepsilon_c'= 11.38$ (compatible with typical semiconductor materials), and vary its imaginary part 
\beq
0<\varepsilon''_c<0.35,
\eeq
the normalized period 
\beq
0.1<a/\lambda_0<0.7,
\eeq
and the normalized cylinder radius
\beq
0<r_c/a<0.5.
\eeq
The above constraints account for the technological feasibility of the required gain level, \cite{Hatori:2000,Nezhad:2004fg,Yu:2008,Geskus:2012} the range of validity of the effective-medium model,\cite{Wu:2006hd} and the geometrical consistency of the unit cell, respectively.
Figure \ref{Figure6}(c) shows, in the 3-D parameter space ($\varepsilon''_c, a/\lambda_0, r_c/a$), a set of possible candidate configurations that satisfy the asymptotic condition in (\ref{eq:boundasmu}) for the existence of an unattenuated bound mode.

%
\begin{figure}
\begin{center}
\includegraphics [width=8.5cm]{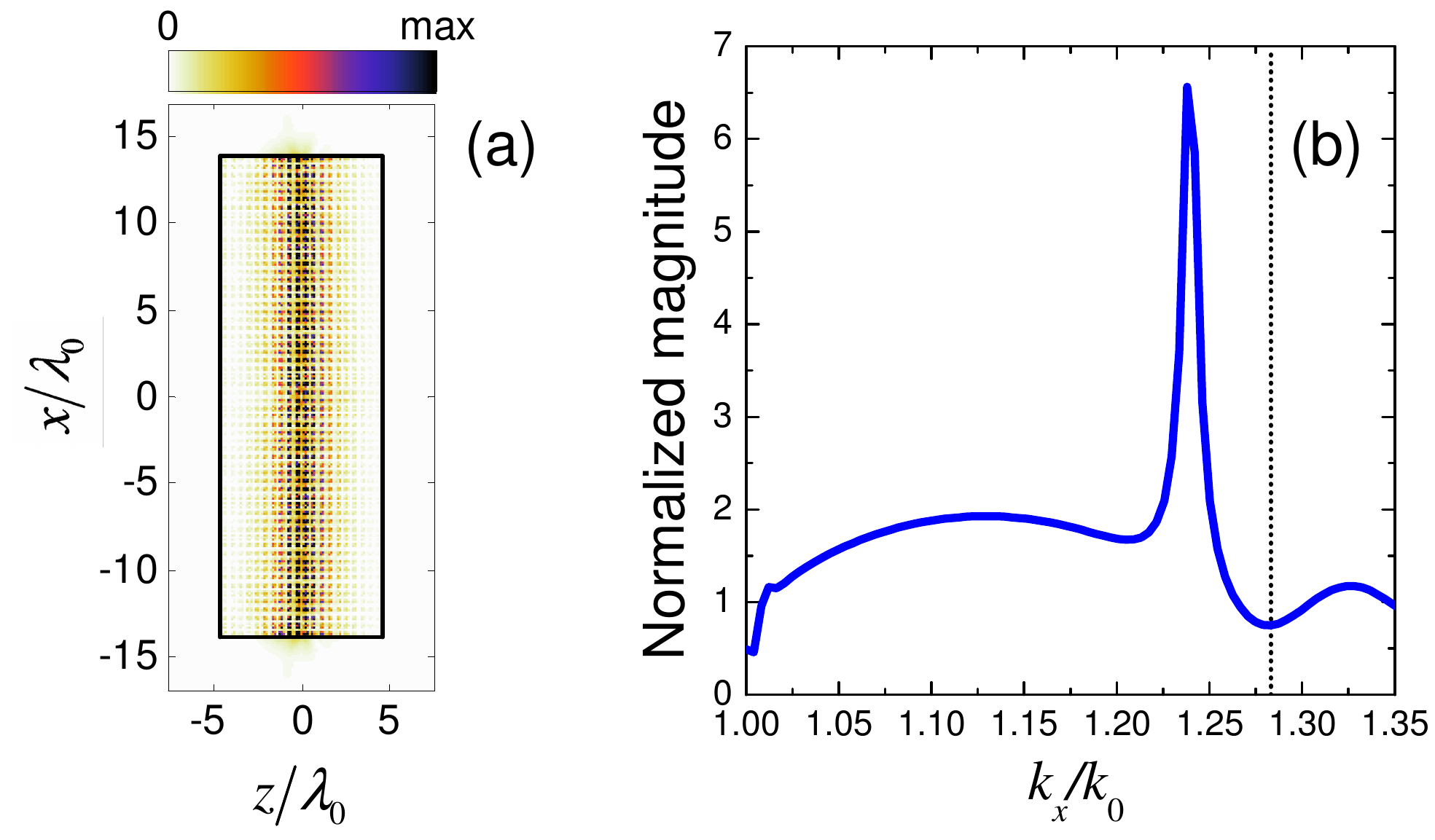}
\end{center}
\caption{(Color online) (a) As in Fig. \ref{Figure4}(b), but for a rod-based metamaterial implementation with $d=4.65\lambda_0$ and finite-size (along $x$) width of $27.9\lambda_0$. Each half of the bi-layer consists of a $10\times 60$ square array of cylindrical rods, with period $a=0.465\lambda_0$, radius $r_c=0.375 a$, and relative permittivity 
$\varepsilon_c=11.38\mp i 0.25$ (for the gain and loss region, respectively). The corresponding effective parameters [cf. (\ref{eq:epse}) and (\ref{eq:mue})] are $\varepsilon_{1e}=0.002-i 0.107$ and $\mu_{1e}=0.567-i0.013$ for $-d<z<0$ (gain), and $\varepsilon^*_{1e}=0.002+i 0.107$ and $\mu^*_{1e}=0.567+i0.013$ for $0<z<d$ (loss).
(b) Field magnitude ($|H_y|$, normalized with respect to the excitation amplitude at a reference plane) at the gain-loss interface $z=0$ for an infinite (along $x$) structure illuminated by an evanescent plane wave, as a function of the $k_x$ wavenumber. Also shown as a reference (black-dotted vertical line) is the theoretical bound-mode propagation constant [cf. (\ref{eq:kxSW1})].}
\label{Figure7}
\end{figure}

%
\begin{figure}
\begin{center}
\includegraphics [width=8.5cm]{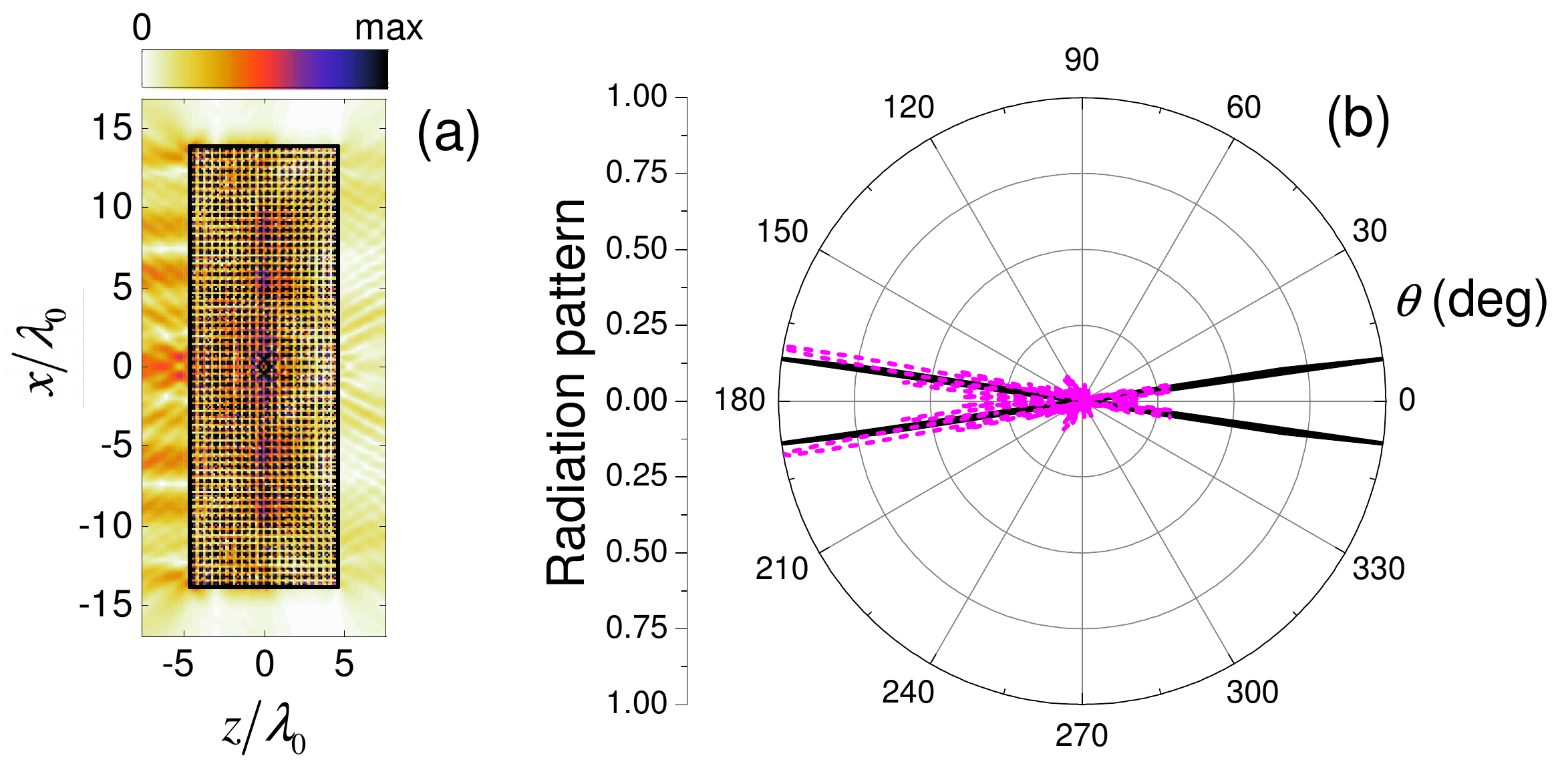}
\end{center}
\caption{(Color online) (a) As in Fig. \ref{Figure7}(a), but for $\varepsilon''_c=\pm 0.05$, i.e., $\varepsilon_e=0.007-i 0.021$ and $\mu_e=0.567-i0.003$. (b) Numerically-computed radiation pattern (with the angle $\theta$ measured with respect to the $z$ axis) compared with leaky-mode-based theoretical prediction in (\ref{eq:LWRP}) for $k_x=(0.147+i 3.4\cdot 10^{-5}) k_0$.}
\label{Figure8}
\end{figure}

\subsection{Results}

As an example, among the possible configurations in Fig. \ref{Figure6}(c), we consider $\varepsilon''_c=0.25$, $a=0.465\lambda_0$, and $r_c=0.375a$, which yields the effective parameters [cf. (\ref{eq:epse}) and (\ref{eq:mue})] 
$\varepsilon_{1e}=0.002-i 0.107$ and $\mu_{1e}=0.567-i0.013$ for the gain region. Accordingly, the lossy region ($\varepsilon^*_{1e}=0.002+i 0.107$ and $\mu^*_{1e}=0.567+i0.013$) can be synthesized by utilizing the same parameters, but $\varepsilon''_c=-0.25$. Assuming an idealized $\mathcal{PT}$-symmetric bi-layer with such effective parameters
 and $d=4.65\lambda_0$, numerical solution of the dispersion equation (\ref{eq:DD}) [with (\ref{eq:kz1mu})] predicts an unattenuated bound mode with $k_x=1.283k_0$.

Figure \ref{Figure7}(a) shows the numerically-computed  field map pertaining to the actual rod-based metamaterial structure excited by a magnetic line source at the gain-loss interface. Also in this case, a bound mode is clearly visible and, although the transverse localization is mostly dictated by the microstructure geometry, we can verify that the propagation constant is in quantitative good agreement with the theoretical predictions. To this aim, we consider an infinite (along $x$) structure illuminated by an evanescent plane wave, and plot in Fig. \ref{Figure7}(b) the (normalized) field magnitude at the interface $z=0$ as a function of the $k_x$ wavenumber. We observe that the response is strongly peaked around $k_x=1.234 k_0$, thereby indicating a phase-matching with a propagation constant that is only $\sim 3\%$ different than the theoretical prediction above. 

As a further confirmation, we decrease the gain/loss level in the rods to $\varepsilon''_c=0.05$, leaving all other parameters unchanged. This yields the effective parameters $\varepsilon_{1e}=0.007-i 0.021$ and $\mu_{1e}=0.567-i0.003$, for which the bound-mode condition in (\ref{eq:boundasmu}) is no longer satisfied. Accordingly, numerical solution of the dispersion equation (\ref{eq:DD}) [with (\ref{eq:kz1mu})] now predicts a leaky mode with $k_x=(0.147+i 3.4\cdot 10^{-5})k_0$.

Figure \ref{Figure8} shows the results pertaining to the actual rod-based structure. In particular, from the field map in Fig. \ref{Figure8}(a) the radiative character of the mode is quite evident. Also in this case, looking at the (far-field) radiation patterns in Fig. \ref{Figure8}(b) we find a good agreement with the theoretical prediction [cf. (\ref{eq:LWRP})].

Overall, the above results indicate that the rod-based metamaterial implementation, based on realistic material constituents, reproduces fairly well the waveguiding mechanism of interest, with good agreement between numerical simulations and theoretical predictions. As previously mentioned, the reliance on material constituents with {\em positive} (real-part) permittivity removes possible ambiguities on the actual nature of the phenomenon, which can thus be clearly attributed to the $\mathcal{PT}$-symmetry.

\section{Conclusions and Perspectives}
\label{Sec:Conclusions}

To sum up, we have shown that ENZ metamaterial bi-layers can support $\mathcal{PT}$-symmetry-induced bound modes at the gain-loss interface. These modes propagate without attenuation provided that the gain/loss level exceeds a critical threshold, and otherwise exhibit a leaky (radiative) character. Starting from the analytical studies and parameterizations, we have designed and simulated  possible rod-based metamaterial implementations. 

Overall, our results indicate that this intriguing $\mathcal{PT}$-symmetry-induced waveguiding mechanism can be observed in the presence of gain/loss levels that are compatible with current technological constraints. This may set the stage for interesting applications to reconfigurable nanophotonic platforms, as well as novel strategies for the design of optical switches and modulators. Besides these potential applications, 
we are currently exploring possible alternative metamaterial implementations, as well as the use of more realistic physical models of gain materials.

\appendix

\section{Details on the asymptotic dispersion relationships (\ref{eq:kSW}) and (\ref{eq:kxSW1})}
\label{Sec:AppA}
Assuming the more general (electric and magnetic) scenario of $\mathcal{PT}$-symmetric half-spaces,
\beq
\varepsilon\left(z\right)=\left\{
\begin{array}{ll}
\varepsilon_1,\hspace{5mm} z<0,\\
\varepsilon_1^*,\hspace{5mm} z>0.
\end{array}
\right.,~~~
\mu\left(z\right)=\left\{
\begin{array}{ll}
\mu_1,\hspace{5mm} z<0,\\
\mu_1^*,\hspace{5mm} z>0,
\end{array}
\right.
\eeq
a modal solution exponentially bound at the gain-loss interface can be written as
\beq
H_y\left(x,z\right)=C\exp\left(i k_x x\right)\left\{
\begin{array}{ll}
\exp\left(i k_{z1} z\right),\hspace{7mm}z<0,\\
\exp\left(ik_{z1}^*z\right),\hspace{7mm}z>0,
\end{array}
\right.
\label{eq:ee1}
\eeq
where $C$ denotes a normalization constant, and the continuity condition at the interface $z=0$ is enforced.
From the relevant Maxwell's curl equation, we then calculate the tangential electric field,
\beq
E_x\left(x,z\right)=\frac{\eta_0}{ik_0\varepsilon\left(z\right)}\frac{\partial H_y}{\partial z}\left(x,z\right),
\label{eq:Ex}
\eeq
where $\eta_0$ denotes the vacuum characteristic impedance. Finally, by enforcing its continuity at the interface $z=0$, we
obtain
\beq
\frac{k_{z1}}{\varepsilon_1}=\frac{k_{z1}^*}{\varepsilon_1^*},
\label{eq:ddisp}
\eeq
from which the dispersion relationship in (\ref{eq:kxSW1}) readily follows by squaring and solving with respect to $k_x$. Note that, as a consequence of the squaring, (\ref{eq:kxSW1}) may yield spurious solutions which do not satisfy (\ref{eq:ddisp}). Hence, the additional constraint (\ref{eq:addcond}) [which derives directly from (\ref{eq:ddisp})] needs to be enforced.

The dispersion relationship in (\ref{eq:kSW}) immediately follows by particularizing (\ref{eq:kxSW1}) to the non-magnetic case $\mu_1=1$.

\section{Details on the dispersion equation (\ref{eq:DD})}
\label{Sec:AppB}
For the $\mathcal{PT}$-symmetric bi-layer in Fig. \ref{Figure1}, a modal solution exponentially bound at the gain-loss interface $z=0$ can be expressed as
\begin{widetext}
\beq
H_y\left(x,z\right)=\exp\left(i k_x x\right)\left\{
\begin{array}{llll}
C_1\exp\left(-i k_{z0} z\right),\hspace{35mm}z<-d,\\
C_2 \exp\left(i k_{z1} z\right) +C_3\exp\left(-i k_{z1} z\right),\hspace{10mm} -d<z<0,\\
C_4 \exp\left(i k_{z1}^* z\right) +C_5\exp\left(-i k_{z1}^* z\right),\hspace{10mm} 0<z<d,\\
C_6 \exp\left(ik_{z0} z\right),\hspace{38mm}z>d,
\end{array}
\right.
\label{eq:ee}
\eeq
\end{widetext}
with $k_{z1}$ and $k_{z0}$ given in (\ref{eq:kz1}) and (\ref{eq:kz0}), respectively, and the unknown expansion coefficients $C_j$, $j=1,...,6$ to be calculated by enforcing the continuity of the magnetic [(\ref{eq:ee})] and electric [cf. (\ref{eq:Ex}) with (\ref{eq:epss})] tangential fields at the three interfaces $z=0$ and $z=\pm d$. This yields a $6\times6$ homogeneous linear system of equations, whose nontrivial solutions can be found by zeroing the system-matrix determinant, viz.,
\begin{eqnarray}
\det&=&\varepsilon_1^*k_{z1}^2\tau_1
\left(
k_{z1}^*-i\varepsilon_1^*k_{z0}\tau_1^*
\right)\nonumber\\
&+&\varepsilon_1^2k_{z0}k_{z1}^*\tau_1
\left(
\varepsilon_1^*k_{z0}-ik_{z1}^*\tau_1^*
\right)\nonumber\\
&+&\varepsilon_1k_{z1}
\left[
2i\varepsilon_1^*k_{z0} k_{z1}^*
+
\left(\varepsilon_1^*\right)^2k_{z0}^2\tau_1^*+\left(
k_{z1}^*
\right)^2 \tau_1^*
\right]\nonumber\\
&=&i k_{z0} \left\{\left|\tau_1\right|^2\mbox{Re}\left[\varepsilon_1^2\left(k_{z1}^*\right)^2\right]
-\left|\varepsilon_1\right|^2 \left|k_{z1}\right|^2
\right\}\nonumber\\
&-&\left|k_{z1}\right|^2\mbox{Re}\left(\varepsilon_1k_{z1}^*\tau_1^*\right)
-\mbox{Re}\left(\left|\varepsilon_1\right|^2\varepsilon_1 k_{z0}^2k_{z1}^*\tau_1\right),
\label{eq:det}
\end{eqnarray}
where the last equality follows from simplifications exploiting the $\mathcal{PT}$-symmetric character.
The dispersion relationship in (\ref{eq:ddisp}) readily follows by zeroing (\ref{eq:det}) and neglecting [in view of the assumed ENZ regime, cf. (\ref{eq:ENZ1})] the third-order term in $\varepsilon_1$.

\section{Details on the leaky-to-bound mode transition}
\label{Sec:AppC}
We now prove that, for gain/loss levels beyond the threshold $\varepsilon''_t$ in (\ref{eq:varepsilonc}) the $\mathcal{PT}$-symmetric ENZ bi-layer in Fig. \ref{Figure1} supports a bound mode [cf. (\ref{eq:BW})] with real propagation constant (i.e., no attenuation).  To this aim, it is expedient to rewrite the dispersion equation (\ref{eq:ddisp}) as
\beq
F\left(k_x\right)=0,
\label{eq:F12}
\eeq
with
\begin{eqnarray}
F\left(k_x\right)&=&-\alpha_0\lbrace|\tau_1|^2\mbox{Re}[\varepsilon_1^2(k_{z1}^*)^2]-|\varepsilon_1|^2|k_{z1}^2|
\rbrace\nonumber\\
&-&|k_{z1}|^2 \mbox{Re}(\varepsilon_1k_{z1}^*\tau_1^*),
\label{eq:dispF}
\end{eqnarray}
and
\beq
\alpha_0=-i k_{z0}=\sqrt{k_x^2-k_0^2}.
\label{eq:gamma0}
\eeq
In such a way, the {\em real} character of the dispersion equation in the parameter range of interest $k_x\ge k_0$ is emphasized, and a simple bracketing strategy can be exploited to prove the existence of real-valued roots.

First we consider the asymptotic limit $k_x\gg k_0$, for which we straightforwardly obtain from (\ref{eq:gamma0}) and (\ref{eq:kz1})
\beq
\bigl.\alpha_0\bigr|_{k_x\gg k_0} \sim k_x,~~
\bigl.k_{z1}\bigr|_{k_x\gg k_0} \sim -i k_x
\label{eq:kz1a}
\eeq
and hence, from (\ref{eq:tau1}),
\beq
\bigl.\tau_1\bigr|_{k_x\gg k_0} \sim -i.
\label{eq:tau1c}
\eeq
By substituting (\ref{eq:kz1a}) and (\ref{eq:tau1c}) in (\ref{eq:dispF}), we then obtain
\begin{eqnarray}
\bigl.F\left(k_x\right)\bigr|_{k_x\gg k_0} &\sim& k_x^3\left[
\varepsilon'+\mbox{Re}\left(\varepsilon_1^2\right)+\left|\varepsilon_1\right|^2
\right]\nonumber\\
&\approx& \varepsilon' k_x^3>0,
\label{eq:FFas}
\end{eqnarray}
where the last approximate equality stems from neglecting [in view of the assumed ENZ regime, cf. (\ref{eq:ENZ1})] second-order terms in $\varepsilon_1$. We have thus shown that the left-hand-side of the dispersion equation (\ref{eq:F12}) is {\em always positive} in the asymptotic limit $k_x\gg k_0$.

Next, we consider the grazing condition $k_x=k_0$, for which  (\ref{eq:gamma0}), (\ref{eq:kz1}) and (\ref{eq:tau1}) yield
\beq
\bigl.\alpha_0\bigr|_{k_x= k_0} =0,~~
\bigl.k_{z1}\bigr|_{k_x=k_0}=k_0\sqrt{\varepsilon_1-1}\approx k_0\left(\frac{\varepsilon_1}{2}-i\right),
\label{eq:kz1b}
\eeq
and
\beq
\bigl.\tau_1\bigr|_{k_x=k_0}\approx -i\tau_0-i\frac{\varepsilon_1 k_0d}{2}\left(\tau_0^2-1\right),
\label{eq:tau1b}
\eeq
respectively, with the approximate equality stemming from first-order McLaurin expansions in $\varepsilon_1$. Substitution of (\ref{eq:kz1b}) and (\ref{eq:tau1b}) in (\ref{eq:dispF}) finally yields
\begin{eqnarray}
F\left(k_0\right)&\approx&-\frac{k^3_0\left|\varepsilon_1-1\right|}{4}\left[
2\tau_0 \left(\left|\varepsilon_1\right|^2-2\varepsilon'\right)
\right.\nonumber\\
&+&
\left.
\left|\varepsilon_1\right|^2 k_0 d\left(\varepsilon'-2\right)
\left(\tau_0^2-1\right)
\right].
\label{eq:N1pi2}
\end{eqnarray}
Recalling the asymptotic behavior in (\ref{eq:FFas}), we can conclude that if $F\left(k_0\right)<0$, the dispersion equation in (\ref{eq:dispF}) must admit a {\em real-valued} solution $k_x\ge k_0$, which corresponds to a bound mode [cf. (\ref{eq:BW})]. By solving (\ref{eq:N1pi2}) with respect to the gain/loss level $\varepsilon''$, this condition can be parameterized as
\beq
\varepsilon''>\varepsilon''_t,
\eeq
with the threshold $\varepsilon''_t$ given in (\ref{eq:varepsilonc}). Moreover, since it can be numerically verified that, within the parameter range of interest, $F(k_x)$ is a monotonic function, the above condition turns out to be not only sufficient, but also necessary.

For subthreshold gain/loss levels, {\em complex-valued} solutions are found instead, which generally exhibit the leaky-mode character in (\ref{eq:LW}).

\section{Details on the numerical simulations}
\label{Sec:AppD}
All the numerical simulations in our study are carried out by means of the finite-element-based commercial software package COMSOL Multiphysics.\cite{COMSOL} 
In particular, we utilize the RF module and the frequency-domain solver.

For the finite-size configurations in Figs. \ref{Figure4}(b), \ref{Figure5}(b), \ref{Figure7}(a) and \ref{Figure8}(a), we utilize a magnetic line-current excitation located at the center of the structure ($x=z=0$), perfectly-matched-layer terminations for the computational domain, and a triangular mesh with adaptive element size. This results in a number of elements on the order of $2.8\cdot 10^5$ and $1.3\cdot 10^6$ for the idealized [cf. Figs. \ref{Figure4}(b) and \ref{Figure5}(b)] and rod-based [cf. \ref{Figure7}(a) and \ref{Figure8}(a)] configurations, respectively. The (far-field) radiation patterns [cf. Figs. \ref{Figure5}(c) and \ref{Figure8}(b)] are straightforwardly obtained by utilizing the post-processing tools in the RF module.\cite{COMSOL}

The results in Fig. \ref{Figure7}(b) refer instead to an infinite (along $x$) structure, simulated by means of a unit-cell consisting of a single row of rods with phase-shift boundary conditions, and excited via a wave-port\cite{COMSOL} by an evanescent plane-wave.


%

\newpage

\end{document}